\documentstyle[12pt]{article}
\catcode`\@=11
\def\marginnote#1{}
\newcount\hour
\newcount\minute
\newtoks\amorpm
\hour=\time\divide\hour by60
\minute=\time{\multiply\hour by60 \global\advance\minute by-\hour}
\edef\standardtime{{\ifnum\hour<12 \global\amorpm={am}%
        \else\global\amorpm={pm}\advance\hour by-12 \fi
        \ifnum\hour=0 \hour=12 \fi
        \number\hour:\ifnum\minute<10 0\fi\number\minute\the\amorpm}}
\edef\militarytime{\number\hour:\ifnum\minute<10 0\fi\number\minute}

\def\draftlabel#1{{\@bsphack\if@filesw {\let\thepage\relax
   \xdef\@gtempa{\write\@auxout{\string
      \newlabel{#1}{{\@currentlabel}{\thepage}}}}}\@gtempa
   \if@nobreak \ifvmode\nobreak\fi\fi\fi\@esphack}
        \gdef\@eqnlabel{#1}}
\def\@eqnlabel{}
\def\@vacuum{}
\def\draftmarginnote#1{\marginpar{\raggedright\scriptsize\tt#1}}
\def\draft{\oddsidemargin -.5truein
        \def\@oddfoot{\sl preliminary draft \hfil
        \rm\thepage\hfil\sl\today\quad\militarytime}
        \let\@evenfoot\@oddfoot \overfullrule 3pt
        \let\label=\draftlabel
        \let\marginnote=\draftmarginnote
@
 \def\@eqnnum{(\theequation)\rlap{\kern\marginparsep\tt\@eqnlabel}%
\global\let\@eqnlabel\@vacuum}  }

\def\numberbysection{\@addtoreset{equation}{section}
        \def\theequation{\thesection.\arabic{equation}}}

\def\underline#1{\relax\ifmmode\@@underline#1\else
        $\@@underline{\hbox{#1}}$\relax\fi}

\catcode`@=12
\relax
\numberbysection
\newtheorem{definition}{\bf Definition}

\newtheorem{lemma}{\bf Lemma}

\def\qed{{\bf Q.E.D. }}
\def\proof{{\sl Proof:  }}
\def\nnn{\hfill \nonumber \\}
\def\Xb{{\bar X}}
\def\Yb{{\bar Y}}
\def\Ab{\bar A}
\def\Bb{\bar B}
\def\Cb{\bar C}
\def\Db{\bar D}

\def\Tb{\bar T}
\def\Ub{{\bar U}}
\def\Yb{{\bar Y}}

\def\zb{\bar z}

\def\bb{\bar b}
\def\eb{\bar e}
\def\fb{\bar f}

\def\chib{\bar \chi}
\def\zetab{{\bar \zeta}}
\def\wb{{\bar w}}

\def\1b{\bar 1}
\def\2b{\bar 2}
\def\nb{\bar n}
\def\ub{\bar u}

\def\Cb{\bar C}
\def\Db{\bar D}

\def\bfX{{\bf X}}
\def\bfY{{\bf Y}}
\def\partialb{\bar \partial}

\def\0b{\bar 0}

\def\Yb{\bar Y}

\def\rhob{\bar{\rho}}

\def\eb{\bar e}

\def\Wr{\hbox{\rm Wr}}

\def\beq{\begin{equation}}
\def\eeq{\end{equation}}
\def\beqa{\begin{eqnarray}}
\def\eeqa{\end{eqnarray}}
\begin{document}
\begin{titlepage}

\nopagebreak
\begin{flushright}

LPTENS--93/38\\
hep-th/9310116
 \\
    October 93
\end{flushright}

\vglue 3  true cm
\begin{center}
{\large \bf

INTRODUCTION TO  DIFFERENTIAL W-GEOMETRY}
\vglue 1.5 true cm
{\bf Jean-Loup~GERVAIS}\\
\medskip
{\footnotesize Laboratoire de Physique Th\'eorique de
l'\'Ecole Normale Sup\'erieure\footnote{Unit\'e Propre du
Centre National de la Recherche Scientifique,
associ\'ee \`a l'\'Ecole Normale Sup\'erieure et \`a
l'Universit\'e
de Paris-Sud.},\\
24 rue Lhomond, 75231 Paris CEDEX 05, ~France}.
\end{center}
\vfill
\begin{abstract}
\baselineskip .4 true cm
\noindent
{\footnotesize
Ideas recently put forward by Y. Matsuo and the author are
summarized on the  example of
the simplest ($W_3$) generalization of two-dimensional gravity.
These notes are based on lectures given at the workshop  `` Strings,
Conformal Models and Topological Field Theories'', Cargese 12-21 May 1993;
and the meeting ``String 93'', Berkeley, 24-29 May 1993.
}
\end{abstract}
\vglue 3 true cm
\vfill
\end{titlepage}

\section{Introduction}

In many ways, W-algebras\index{W algebra}
are natural generalizations of the Virasoro
algebra\index{Virasoro algebra}.
 They were first introduced as  consistent operator-algebras
 involving
operators of spins higher than two\cite{Z}. Moreover, the Virasoro
algebra is intrinsically related with the Liouville  theory which
is
the Toda theory associated with the Lie algebra\index{Lie algebra}
 $A_1$, and this
relationship extends to W-algebras which  are in correspondence with
the family of conformal Toda systems associated with arbitrary simple
Lie algebras\cite{BG1}. Another point is that the deep connection
between
Virasoro algebra and KdV hierarchy has  a natural extention\cite{GY}
  to  W-algebras
and higher KdV\index{KdV equation} (KP) hierachies\cite{Sato}.
On the other hand, W symmetries
exhibit strikingly novel features. First,
they  are basically non-linear algebras. Since the transformation
laws  of primary fields  contain higher derivatives,  product of primaries
 are not primaries at the classical level. Naive tensor-products of
commuting representations do not form representations. A related novel
feature is that W-algebras generalize the diffeomorphisms of the circle
by including derivatives of degree higher than one. Going beyond linear
approximation (tangent space ) is  a highly non-trivial step.
Taking higher order derivatives changes the shape of the world-sheet
in the target-space, thus W-geometry should be related to the extrinsic
geometry of the embedding.
Finally, Virasoro algebras are notoriously related to
Riemann surfaces\index{Riemann surfaces}.
The W-generalization of the latter notion is a fascinating problem.

In a series of recent papers, we have developed a geometrical framework
 for the   class of
Conformal field Theory CFT\index{Conformal field Theory CFT}
mentioned above,
where these features emerge from the standard Riemannian geometry
of particular manifolds which we called
$W$ surfaces\cite{GM1}--\cite{GM3}. These references cover quite a
lot of material, and the present lecture will go in the opposite way.
Leaving the description of the general scheme to
refs\cite{GM1}--\cite{GM3}, we shall, instead,  illustrate the ideas  by
two explicit examples: the 2D gravity case (section 2), and its simplest
generalization to the $W_3$ gravity
 (section 3). In both cases, we
discuss the two current approaches, namely, the conformal one where
W-gravity is identified with a conformal Toda
(or Liouville theory\index{Liouville theory} for
2D gravity)  theory, and the
light-cone approach.

Before starting, however,
let us recall how our general scheme goes. One  basic point is
to make use of the fact that one deals with integrable
models\index{Integrable models}, but geometrically
 there are two aspects.
The first uses extrinsic geometry\index{extrinsic geometry}.
In ref.\cite{GM2}, we showed
  that the
$A_n$-W--geometry corresponds  to the embedding of holomorphic
two-dimensional surfaces in the complex
projective space\index{complex projective space}
$CP^n$. These (W) surfaces are
specified by
embedding equations  of the form $X^A=f^A(z)$,
$\Xb^{\Ab}=\fb^{\Ab}(\zb)$,
where $z$, and $\zb$ are the two surface-parameters. The fact that
they are functions of a single  variable is equivalent to
the Toda field-equations, so that this describes W gravity in
the conformal gauge.
These functions have a natural extension to $CP^n$ using the
higher variables $z^{k}$, $\zb^{k}$
 of the Toda hierarchy\index{Toda hierarchy}  of integrable flows, and this
provides a local parametrization of $CP^n$. The original variables
$z$ and $\zb$ are identified with $z^{1}$, $\zb^{1}$, respectively.
 For  the embedding
 functions the extension is such that they become functions of
half  of the variables noted
$f^A([z])=f^A(z^{0}, \cdots z^{n})$, and
$\fb^{\Ab}([\zb])=\fb^{\Ab}( \zb^{0}, \cdots \zb^{n})$ such that
\beq
{\partial f^A([z]) \over \partial z^{k}}  =
{\partial^k f^A([z]) \over (\partial z)^k}, \quad  \quad
{\partial \fb^{\Ab}([\zb]) \over \partial \zb^{k}}  =
{\partial^k \fb^{\Ab}([\zb]) \over (\partial \zb)^k}
\label{1.1}
\eeq
 One  main virtue of the coordinates
$z^{k}$, $\zb^{k}$ is that,  due to the last equations,
 higher derivaties in $z$ and $\zb$ are changed
to first-order ones, and this is how our geometrical
scheme gets rid of the  troublesome higher derivatives of the
usual approachs.

The second aspect\cite{GM1}\cite{GM2}
 only makes use of intrinsic geometries, but introduces
a family of associated surfaces in the standard Grassmannians associated
with $CP^n$. This is useful to discuss global aspects by using the
fact developed in ref.\cite{GM2}, that W surfaces are instantons of
non-linear $\sigma$ models.  We shall not dwell into this aspect in
these lectures.

 So far this is only for the conformal gauge.
Concerning the light-cone approach, our recent insight\cite{GM3}
 is that,  for
any K\"ahler manifold,  there exist  changes   of coordinates such that
 metric tensors
  of the light-cone type come out. This allows us to relate
conformal and light-cone descriptions  by   diffeomorphisms.

Finally, let us stress that we shall remain entirely at the level
of classical field theory, thus describing only the $C\to \infty $ limit
of the problem. The quantum approach to Toda theories is
making steady progress\cite{G1}--\cite{GS}, \cite{CGR1}--\cite{GS2}
 but its connection with
the present geometrical scheme remains as a fascinating problem for the
future.

\section{Two-dimensional Gravity}

In this part we discuss the case of two-dimensional gravity
in some details, using methods that will be later on generalized to
W gravity.
With general world-sheet parameters $\xi$, the Weyl anomaly takes the
form
\beq
S={1\over \gamma} \int d_2\xi ( {1\over 2}{\cal R} {1\over \Delta }
{\cal R} +
\mu \sqrt {
-{\cal G}})
\label{2.1}
\eeq
In this expression, $\Delta$ is the Laplacian with the 2D-gravity metric
$ {\cal G}_{\alpha, \, \beta}$,
$\gamma$ is   the coupling constant, ${\cal R}$
the scalar curvature, and $\mu$
the cosmological constant.
\subsection{Conformal gauge}

First,  choose conformal coordinates $z$ and
$\zb$, so that the arc length takes the
form\footnote{This may not be possible globally, but we shall
only consider the local aspects}
\beq
ds^2=2e^{2 \Phi} dz d\zb,
\label{2.2}
\eeq
the Weyl anomaly becomes the Liouville action
\beq
S={1 \over \gamma} \int d_2 z  ( -{1\over 2}
\partial \Phi \partialb \Phi  + \mu e^{2 \Phi}
).
\label{2.3}
\eeq
In general $\partial$ and $\partialb$ denote $\partial /\partial z$, and
$\partial /\partial \zb$ respectively. At the level of the
present discussion we may always set $\mu =1$ by a shift of $\phi$, and we
shall do so from now on. The general solution of the Liouville
equation is given by  the holomorphic decomposition
\beq
e^{- \Phi} =\sum_{j=1}^2 \chi_j(z) \chib_j(\zb).
\label{2.4}
\eeq
The functions $\chi_j(z)$, and $\chib_j(\zb)$ are pairs of solutions
of the differential equations
\beq
-{d^2 \chi_j(z) \over (dz)^2} +T(z)\chi_j(z) =0, \quad
-{d^2 \chib_j(\zb) \over (d\zb)^2} +\Tb(\zb)\chib_j(\zb) =0,
\label{2.5}
\eeq
where $T$ and $\Tb$ are the two non-vanishing components of the
stress-energy tensor. They are normalized so that
\beq
\chi_1(z) {d \chi_2(z) \over dz}-
\chi_2(z){d \chi_1(z) \over dz}=
\chib_1(\zb) {d \chib_2(\zb) \over d\zb}-
\chi_2(\zb){d \chi_1(\zb) \over d\zb}=1.
\label{2.6}
\eeq
At this point, and in the following,  we need a
simple mathematical lemma concerning differential equations, which we
state once for all next.
\begin{lemma}{Automatic differential equation. }

Consider $N$ functions $f^A$ of a single variable $x$, whose
Wronskian
does not vanish.
They satisfy a differential equation of the form
\beq
f^{(N)A}(x)=\sum_{\ell=1}^N \kappa_\ell(x)
f^{(N-\ell)A}(x).
\label{2.7}
\eeq
\end{lemma}

\proof Recall the definition of the Wronskian:
\beq
\Wr(f^1,\cdots ,\, f^N)\equiv
\left | \begin{array}{ccc}
f^{1}  & \cdots  & f^{N} \\
f^{(1)\, 1}  & \cdots  & f^{(1)\, N} \\
\vdots     & \cdots         & \vdots  \\
f^{(N-1)\, 1}  & \cdots  & f^{(N-1)\, N} \\
\end{array}
\right |.
\label{2.8}
\eeq
In this formula, and hereafter,
upper indices in between parentheses denote the order of derivatives.
Clearly, any of the functions $f^A$ may be trivially written as
a linear combination of the type $\sum_B c^A_B f^B$. It is then clear that
\[
\left | \begin{array}{cccc}
f^{1}  & \cdots  & f^{N} & f  \\
f^{(1)\, 1}  & \cdots  & f^{(1)\, N} & f^{(1)} \\
\vdots     & \cdots         & \vdots  \\
f^{(N)\, 1}  & \cdots  & f^{(N)\, N} & f^{(N)} \\
\end{array}
\right | =0.
\]
Expanding this determinant with respect to the last column immediately
gives Eq.\ref{2.7}. \qed   Moreover, one immediately sees that
\beq
\kappa_1={d \ln [\Wr(f^1,\cdots ,\, f^n)]\over dx}
\label{2.9}
\eeq

Returning to our main line, we see that, in the differential equations
Eq.\ref{2.5} the first order term vanishes, so that the Wronskians
$\Wr(\chi)$ and $\Wr(\chib)$ are constant, and may be chosen equal to
one (see  Eq.\ref{2.6}). For our geometrical description this is not
appropriate, however. The basic reason is as follows. Under conformal
transformations $\delta z=\epsilon(z)$, the $\chi_j$
fields\footnote{In the conformal gauge $\chi$ and $\chib$ are on the
same footing. When we discuss properties of chiral
component we some times
talk about the $\chi$ fields as an example. Clearly,
the $\chib$ fields are analogous.}  transform as
primary fields of weight $-1/2$ (that is such that
$\chi_j(z)\, (dz)^{-1/2}$
is invariant). This is consistent, since it follows that the Wronskian
$\Wr(\chi)$ transforms  with weight zero, so that condition Eq.\ref{2.6}
is conformally invariant. In our geometrical description, these functions
will become geometrical objects whose form should not change
under conformal transformation, so that they should transform with weight
zero.  We may change the conformal weights
by using a projective description.  For
this we define
\beq
f^A=\sqrt{w(z)}\>  \chi_{A+1}, \quad
\fb^{A}=\sqrt{\wb(\zb)}\>  \chib_{A+1}, \quad A=0,\, 1,
\label{2.10}
\eeq
so that
\beq
\Wr(f^0,\, f^1)=w(z),\quad \Wr(\fb^0,\, \fb^1)=\wb(z)
\label{2.11}
\eeq
where  $w$ and $\wb$ are arbitary functions of a single variable.
Now the $f$'s  satisfy a more general differential equation
of the type Eq.\ref{2.7}, where $\kappa_1\ne 0$. Substituting into
Eq.\ref{2.4}, we derive the arc length
\beq
ds^2=dz d\zb {\Wr(f^0, f^1) \Wr(\fb^0, \fb^1) \over
(f^0\fb^0+f^1\fb^1)^2 },
\label{2.12}
\eeq
and, re-arranging the terms, we may write
\[
ds^2=dz d\zb
\left ({ -(f^{(1)0}\fb^{(1)0}+f^{(1)1}\fb^{(1)1})
(f^{0}\fb^{0}+f^{1}\fb^{1})\over (f^{0}\fb^{0}+f^{1}\fb^{1})^2} +
\right.
\]
\beq
\left. {(f^{(1)0}\fb^{0}+f^{(1)1}\fb^{1})
(f^{0}\fb^{(1)0}+f^{1}\fb^{(1)1})
 \over
(f^{0}\fb^{0}+f^{1}\fb^{1})^2}\right ).
\label{2.13}
\eeq
As is well-known $\Wr(\rho f^0, \rho f^1) =\rho^2 \Wr(f^0,  f^1)$, for
arbitrary function $\rho$, so that\footnote{this is
obviously true by construction.}
 Eq.\ref{2.12}  is invariant under
$f^A\to \rho f^A$, and $\fb^A\to \rhob \fb^A$.  We arrive
 at  a projective
structure,
 following our general scheme\cite{GM2} where $A_n$-$W$ geometries are
described in  $CP^n$  (we will have $n=1$ in the present case).
Let us recall some useful definitions at this point.

\begin{definition}{K\"ahler manifolds}

A K\"ahler manifold of real dimension $2n$  is  complex manifold
with a special  class of coordinates
 $X^A$, $X^{\Ab}$,
$1\leq A,\, \Ab \leq  n$, such that the only components of the
metric are $G_{A \, \Bb}=G_{\Bb\, A}$, and
\beq
 G_{A \, \Bb}=\partial_A \partial_{\Bb} K,
\label{2.14}
\eeq
 where
$K$ is the K\"ahler potential\index{Kahler potential}.
\end{definition}
In general we denote the
differential operators $\partial/\partial X^A$, and
 $\partial/\partial X^{\Bb}$ by $\partial_A$, and  $\partial_{\Bb}$.

\begin{definition}{Complex projective space $CP^n$}

The complex projective space $CP^n$ is defined   from  the
 trivial  complex space $C^{n+1}$ with coordinates
$X^A$, $\Xb^{\Ab}$, $ A, \Ab =0, \cdots n$,  by identifying any
two points $X, \Xb$, and $Y$, $\Yb$
related by the scale transformation
\beq
 X^A = Y^A\rho, \quad
 {\rm and }
\quad \Xb^{\Ab} = \bar{Y}^{\Ab}\rhob.
\label{2.15}
\eeq
\end{definition}

This is a simplest non-trivial   example of a K\"ahler manifold.
The metric on this space is given by the Fubini-Study equation
\beq
G_{A\Ab} = \left (\delta_{A\Ab}\left (\sum_{B = 0}^n X^B\Xb^B
\right )
   - X^{\Ab}\Xb^{A}\right )\Bigl /(\sum_{B = 0}^n X^B\Xb^B)^2,
\label{2.16}
\eeq
whose K\"ahler potential is given by
\beq
 K = \ln\sum_{A = 0}^n X^A\Xb^{\Ab}.
\label{2.17}
\eeq
Eq.\ref{2.16} is invariant under the scale transformation Eq.\ref{2.15}.
Using this freedom, it is customary to parametrize $CP^n$, by letting
one component (say, $X^0$) equal 1.

Closing the parenthesis, we return to Eq.\ref{2.13}. Consider
the space $CP^1$. The equations
\beq
X^A=e^{\zeta} f^A(z), \quad \Xb^A=e^{\zetab} \fb^A(\zb),
\quad A=0,\, 1,
\label{2.18}
\eeq
define a holomorphic change of coordinates in $C^2$
 from $X^A$ to $\zeta, z$.
It is easy to see that Eq.\ref{2.13} is equivalent to
\beq
G_{z\zb}\equiv e^{2\Phi}= f^{(1)A} \fb^{(1)\Bb}
G_{A\, \Bb} \Bigl |_{X^A=f^A,\, \Xb^{\Ab}=\fb^{\Ab}}
\label{2.19}
\eeq
By construction, Eq.\ref{2.13} is
invariant under the rescaling $f\to \rho(z) f$, $\fb \to \rhob(\zb) \fb$.
Thus we may let $\zeta=\zetab=0$, and  the geometrical meaning of
 Eqs.\ref{2.18}, Eq.\ref{2.19} is that  $z$ $\zb$ are  parameters of
$CP^1$, such that the Liouville exponential $G_{z,\zb}$ is equivalent
to the metric tensor of Fubini-Study.  Thus any solution of Liouville
equation defines a  local holomorphic parametrization of $CP^1$.
Note that, since  the change of coordinate is holomorphic,
the K\"ahler
condition is preserved. Indeed, one has
\beq
G_{z\zb} =\partial \partialb {\cal K}, \quad
{\cal K}= \ln (f^0\fb^0+f^1\fb^1).
\label{2.20}
\eeq
%====================================================
\subsection{Light-cone gauge}
\subsubsection{The Weyl anomaly}
First, rederive the ideas of ref.\cite{P1} is the present context.
At the level of Eq.\ref{2.1}, one changes coordinates from $z$ $\zb$
to $u$ $\ub$, such that
\beq
e^{2\Phi}dz d\zb=du d\ub +h du^2
\label{2.21}
\eeq
The Weyl anomaly Eq.\ref{2.1} becomes
\beq
S={1\over 2\gamma} \int d_2 u \left ( (\Db^2 h)
{1 \over D \Db -\Db h \Db }
(\Db^2 h) + 2 \right ),
\label{2.22}
\eeq
where $D$ and $\Db$ stand for $\partial /\partial u$, and
$\partial /\partial \ub$, respectively.
Here  the second term is a constant and can be forgotten.
Using the obvious relation
\[
{1 \over D \Db -\Db h \Db } =
{1 \over D  - h \Db }\,{1 \over  \Db }
\]
we may write  after partially integrating
\beq
S={1\over  \pi } \int d_2 u \> h\>  {\cal T},\quad
{\cal T}\equiv  {\pi\over 2\gamma} \Db^2
{1 \over D  - h \Db }
\Db h
\label{2.23}
\eeq
Starting from this expression, we compute
\[
\left ( D  -h \Db -2 (\Db h) \right ) {\cal T}
= {\pi\over 2\gamma}
\Db^2 { D - h \Db
\over D - h \Db} \Db h
\]
so that we get
\beq
\left ( D-  h \Db -2 (\Db h) \right ) {\cal T}
={\pi\over 2\gamma}
\Db^3 h
\label{2.24}
\eeq
Thus the deformed conservation law  are satisfied.
Minimizing $S$, under the form Eq.\ref{2.23},   with respect to
$h$, one finds
\[0={\pi\over 2\gamma} \Db^2
{1 \over D  - h \Db }
\Db h
={\cal T}
\]
so that Polyakov's anomaly equation\index{Polyakov string theory}
 comes out:
\beq
\Db^3 h=0
\label{2.25}
\eeq
\subsubsection{Liouville dynamics}
Next, following ref.\cite{GM3},
 we connect  the results just recalled with the
conformal-gauge Liouville appproach recalled above.
Since the Liouville equations also follow from minimizing $S$, they are
equivalent to Polyakov's anomaly equations. This may be seen as follows.
 Making use of Eq.\ref{2.21}, one
sees that
\beq
h=-{\partial \ub \over \partial z},\quad
e^{2\Phi}= {\partial \ub \over \partial \zb},
\label{2.26}
\eeq
The last equation is immediately solved, thanks to Eq.\ref{2.20}, and
we find that the change of variables form conformal to light-cone
explicitly reads
\beq
u=z,\quad \ub=\partial K = {f^{(1)0}\fb^{0}+f^{(1)1}\fb^{1}\over
f^{0}\fb^{0}+f^{1}\fb^{1}}
\label{2.27}
\eeq
Computing $h$ in terms of $f^1$, and $\fb^1$ one finds.
\[
h=-{f^{(2)0}\fb^{0}+f^{(2)1}\fb^{1}\over
f^{0}\fb^{0}+f^{1}\fb^{1}}
+\left  [ {f^{(1)0}\fb^{0}+f^{(1)1}\fb^{1}\over
f^{0}\fb^{0}+f^{1}\fb^{1}}  \right ]^2
\]
The first term is retransformed by using
the lemma which gives
$f^{(2)A}=\kappa_1 f^{(1)A}+\kappa_0f^{A}$.
One gets
\beq
h= -\kappa_0-\kappa_1 \ub + \ub^2
\label{2.28}
\eeq
For fixed $z=u$, it is  quadratic in $\ub$, so that Eq.\ref{2.25}
is indeed verified.   In terms of the $sl(2,R)$ light-cone currents
we have $J^+=1$, $J^0=\kappa_1/2$, and $J^-=-\kappa_0$.
%================================================================
\section{$W_3$ gravity}
Unfortunately no formula similar to the Weyl-anomaly term Eq.\ref{2.1}
is known at present. Our starting point will be the generalization of the
Liouville equation known as the $A_2$ Toda equation, where the $W_3$
algebra is realized by Poisson bracket, and which should thus describe
$W_3$ gravity in the conformal gauge.

\subsection{Conformal gauge}
\subsubsection{ $W$ surface}
In the $A_2$ Toda theory there are two fields $\phi_1$, $\phi_2$. The
field equations read
\beq
\partial \partialb \phi_1=-e^{2\phi_1-\phi_2}, \qquad
\partial \partialb \phi_2=-e^{2\phi_2-\phi_1}
\label{3.1}
\eeq
with solution
\beqa
e^{- \phi_1} =\sum_{j=1}^3 \chi_j(z) \chib_j(\zb), \qquad
e^{- \phi_2} =\sum_{i<j=1}^3 \chi_{ij}(z) \chib_{ij}(\zb)
\label{3.2}
\eeqa
There are now three pairs of arbitary functions
of a single variable   $\chi_j$, $\chib_j$, the functions
$\chi_{ij}$, and $\chib_{ij}$ are defined by
\beq
\chi_{ij}=\Wr( \chi_i, \, \chi_j),\quad
\chib_{ij}=\Wr( \chib_i, \, \chib_j).
\label{3.3}
\eeq
The Toda equations \ref{3.1} are satisfied if
\beq
\Wr( \chi_1, \, \chi_2, \, \chi_2)=1,\quad
\Wr( \chib_1, \, \chib_2, \, \chib_2)=1,
\label{3.4}
\eeq
One sees that the generalization of the Liouville case is very direct.
The lemma, combined with the last Wronskian condition shows that
the $\chi_j$ satisfy a differential equation of the form
\beq
\chi^{(3)}_j=T\chi^{(1)}_j +W \chi^{}_j
\label{3.5}
\eeq
The potentials are the generators of the $W$ transformations: $T$ is the
stress-energy tensor, and $W$ the $W_3$ charge. Again we relax the
condition \ref{3.4} by letting
\beq
f^A=(w(z))^{1/3}\>  \chi_{A+1}, \quad
\fb^{A}=(\wb(\zb))^{1/3}\>  \chib_{A+1}, \quad A=0,\, 1, \, 2
\label{3.6}
\eeq
so that
\beq
\Wr(f^0,\, f^1, \, f^2)=w(z),\quad \Wr(\fb^0,\, \fb^1, \, \fb^2)=\wb(z)
\label{3.7}
\eeq
\beqa
e^{- \phi_1}& =&{\sum_{A=0}^2 f^A(z) \fb^A(\zb)\over
\left [\Wr(f^0,\, f^1, \, f^2)
\Wr(\fb^0,\, \fb^1, \, \fb^2)\right ]^{1/3}}\nnn
e^{- \phi_2}& =&{\sum_{A<B=0}^2 \Wr(f^A,\, f^B)
\Wr(\fb^A,\, \fb^B)
\over \left [\Wr(f^0,\, f^1, \, f^2)
\Wr(\fb^0,\, \fb^1, \, \fb^2)\right ]^{2/3}}
\label{3.8}
\eeqa
Our task is now to give a meaning for these formulae in Riemannian
geometry. There are two new features as compared with the Liouville
case. First we now have three pairs  of functions, but we still only have
two variables $z$ and $\zb$. Second, derivatives of second degree
appear so that one does not have a direct
tangent-space interpretation. These two
problems are solved simultaneously according to our general scheme.
First, we introduce $CP^2$ as  target space where the equations
\beq
X^A= f^A(z), \quad \Xb^A= \fb^A(\zb),
\quad A=0,\, 1, \, 2,
\label{3.9}
\eeq
define a holomorphic ($W$) surface.  Second, we will also
consider the extrinsic
geometry of these  surfaces  where higher derivatives naturally appear
when one defines normals.
It is convenient to define
\beq
{\widetilde G}_{a \bb} =
\sum_{A,\Bb} f^{(a)A}(z) \fb^{(\bb)\Bb}(\zb)  G_{A \Bb}
\Bigr |_{ X^A=f^A,\, \Xb^{\Ab}=\fb^{\Ab}}={\widetilde G}_{ \bb a}
\label{3.10}
\eeq
Consider first the intrinsic geometry. It is straightforward
to see that,  according to Eq.\ref{3.8}, the arc-length on the surface is
given by
\beq
ds^2\equiv {\widetilde G}_{1 1}dz d\zb =e^{2\phi_1-\phi_2} dz d\zb
\label{3.11}
\eeq
so that this particular exponential of the Toda fields gives the induced
metric on the $W$ surface. The complete geometrical interpretation may be
obtained following the usual method of differential geometry, namely by
considering normals to the surface. There are two of them, noted $e_{1}$,
and $e_{\1b}$. Their components
may be written compactly
 by the generalized Frenet-Serret formulae
\begin{equation}
\label{3.12}
e^A_1 = {1\over \sqrt{{\widetilde G}_{1 \1b} \Delta_{2}}}
\left | \begin{array}{cc}
{\widetilde G}_{1 \1b} & {\widetilde G}_{2 \1b} \\
f^{(1)A} &f^{(2)A}
\end{array}
\right |,\quad   e^{\Ab}_1=0,
\end{equation}
\begin{equation}
\label{3.13}
e^A_{\1b}=0, \quad
e^{\Ab}_{\1b} = {1\over \sqrt{{\widetilde G}_{1 \1b} \Delta_{2}}}
\left | \begin{array}{cc}
{\widetilde G}_{1 \1b} & {\widetilde G}_{1 \2b} \\
\fb^{(1)\Ab} &\fb^{(2)\Ab}
\end{array}
\right |.
\end{equation}
where
\begin{equation}
\label{3.14}
\Delta_2\equiv \left | \begin{array}{cc}
{\widetilde G}_{1 \1b} &  {\widetilde G}_{1 \2b} \\
{\widetilde G}_{2 \1b}   & {\widetilde G}_{2 \2b}
\end{array}\right |
\end{equation}
It is easy to see that,  if we denote by $( \bfX, \bfY)$ the inner  product
$G_{A \Bb} (X^A Y^{\Bb} +Y^A X^{\Bb})$ in $CP^2$, we have
$( e_1, f^{(1)})=0$, $( e_1, \fb^{(1)})=0$,
$( \eb_1, f^{(1)})=0$, $( \eb_1, \fb^{(1)})=0$,
$( e_1, e_1)=( e_{\1b}, e_{\1b})=0$,
$(e_{1}, e_{\1b})=1 $. So that the last equations
 define orthonormalized normals. With
the present notations, the second fundamental form $Q$
 is obtained by writing
the equivalent expressions
\beq
f^{(2)A} -f^{(1)A} {{\widetilde G}_{2 \1b}\over {\widetilde G}_{1 \1b}}
=e^A_1 Q, \quad
\fb^{(2)\Ab} -\fb^{(1)\Ab}
{{\widetilde G}_{2 \1b}\over {\widetilde G}_{1 \1b}}=
e^{\Ab}_{\1b} Q
\label{3.15}
\eeq
with $Q= \sqrt{ \Delta_{2}/{\widetilde G}_{1 \1b}}$.
By explicit computation\footnote{this is done more powerfully
using the fermionic method as explained in ref.\cite{GM2}}
 one finds that $\Delta_2=\exp(3\phi_1)$, so that
\beq
Q=e^{(\phi_1+\phi_2)/2}=\sqrt { \Wr(f^0,\, f^1, \, f^2)
\Wr(\fb^0,\, \fb^1, \, \fb^2)\over
\sum_{A=0}^2 f^A(z) \fb^A(\zb) \sum_{C<B=0}^2 \Wr(f^C,\, f^B)
\Wr(\fb^C,\, \fb^B)}.
\label{3.16}
\eeq
\subsubsection{Associated holomorphic change of coordinates}
Although the geometrical meaning of the Toda fields is now clear,
the description is not covariant under change of parametrization
of $CP^2$, since it involves second derivatives. This problem is solved
by using the fact that the Toda field equations are a subsystem of the
so-called Toda hierarchy of integrable flows that makes use of $6$
real parameters. These parameters provide a parametrization of $C^3$
where the formulae we summarized may be
re-expressed in terms of first order
derivatives only, so that their covariance properties become clear.
This part is best  handled by means of the fermionic method
and Hirota equations\cite{Sato} \cite{GM1} \cite{GM2}. We shall nevertheless
remain at the same
elementary simple-minded level for illustration.
 In terms of
$f$ and $\fb$, the Toda hierarchy equations become extremely simple
(see Eq.\ref{1.1}). The variables are\footnote{
Upper indices are  always  covariant indices. When we want to indicate
powers such as $z$ square, we put parentheses, and write  $(z)^2$.}
 $z^0$, $z^1\equiv z$, $z^2$ denoted
collectively as $[z]$; and
$\zb^0$, $\zb^1\equiv \zb$, $\zb^2$ denoted
collectively as $[\zb]$. The functions $f^A(z)$ and $\fb^{\Ab}(\zb)$
are extended as holomorphic or antiholomorphic functions
$f^A([z])$ and $\fb^{\Ab}([\zb])$ solutions of the equations
\beqa
\partial_0f^A([z])&=f^A([z]),\quad
\partial_2f^A([z])&=(\partial_1)^2 f^A([z])   \nnn
\partialb_0\fb^{\Ab} ([\zb])&=\fb^{\Ab}([\zb]),\quad
\partialb_2\fb^{\Ab} ([\zb])&=(\partialb_1)^2 \fb^{\Ab}([z])
\label{3.17}
\eeqa
with the initial conditions
\beq
f^A([z])=f^A(z),\quad
\fb^{\Ab}(\zb)=\fb^{\Ab}([\zb]), \quad
\hbox{for}\>
z^0=z^2=\zb^0=\zb^2=0
\label{3.18}
\eeq
Of course the differential equations in $z^0$ and $\zb^0$ are trivial
and give simple factors $\exp(z^0)$ or
$\exp(\zb^0)$. Since we work homogeneously we may
forget these variables  completely and keep $z^0$ and $\zb^0$
 equal to zero. The point of the differential equations in $z^2$ is
to turn second derivatives in $z$ into a first order derivatives.
Thus the matrix of inner product of derivatives Eq.\ref{3.10}
is extended to $CP^2$ as
\beq
G([z],\,[\zb])_{a \bb} =
\sum_{A,\Bb} \partial_af^{A}([z])
\partial_{\bb}\fb^{\Bb}([\zb])  G_{A \Bb}
\Bigr |_{ X^A=f^A([z]),\, \Xb^{\Ab}=\fb^{\Ab}([\zb])}
\label{3.19}
\eeq
where  now {\bf only first order derivatives appear}. As a matter of fact,
this last equation is but the Riemannian metric obtained by making the
holomorphic change of variables $X^A=f^A(z,z^2)$,
$\Xb^{\Ab}=\fb^{\Ab}(\zb, \zb^2)$. Thus again, we associate, with any
solution of the Toda equation, a conformal parametrization of the complex
projective space.
Note again that, since the change is conformal, the K\"ahler
condition is preserved. The metric $G([z],\,[\zb])_{a \bb}$
may be written as
\beq
G([z],\,[\zb])_{a \bb}= \partial_a \partial_{\bb}
\ln \left [ \sum_{A} f^{A}([z]) \fb^{A}([\zb]) \right ]
\label{KK}
\eeq

Our next topic is the W transformations of $f$. We shall use the
parametrization such that $X^0=\Xb^0=1$, so that
$f^0([z])=\fb^{0}([\zb])=1$. Due to these  conditions, and
according to the lemma, the functions $f^A(z)$ are solution of a
differential equation of the form
\beq
f^{(3)A}(z)=\lambda(z)f^{(2)A}(z) +\mu(z) f^{(1)A}(z)
\label{3.20}
\eeq
there is no term without derivative since $f^0=1$ is a solution. The most
general  infinitesimal transformation that leaves the condition
$f^0=1$ invariant is
\beq
\delta f^A(z)=(\alpha(z)\partial+\beta(z)\partial^2) f^A(z), \quad
\label{3.21}
\eeq
Higher derivatives may be re-expressed in terms of the first two
by means of Eq.\ref{3.20}. Consider, now the extension to
$CP^2$.
  To simplify
the notation we  will let $t=z^2$, so that, according to
 Eqs.\ref{3.17},
\beq
\partial_t f^A(z,t)=(\partial)^2 f^A(z,t).
\label{3.22}
\eeq
 This is used in practice  as follows. Consider
\[
\delta_n f(z) = \epsilon(z) (\partial)^n f(z)
\]
Clearly one has
\beq
f^A(z,t) =e^{t (\partial)^2} f^A(z,0)
\label{3.23}
\eeq
One writes
\[
\delta_n f(z,t)= e^{t (\partial)^2} \delta f(z) =
\epsilon(z+2t \partial ) e^{t (\partial)^2} \partial^n f(z)
=\epsilon(z+2t \partial ) (\partial)^n f(z,t)
\]
where $\epsilon(z+2t \partial )$ is defined by
\beq
\epsilon(z+2t \partial )=\sum_r {\epsilon_r \over r !}
(z+2t \partial )^r, \quad
\epsilon_r =(\partial)^r\epsilon(z)|_{z=0}
\label{3.24}
\eeq
In general, one sees that an arbitrary $W$ transformation leads to
\beq
\delta f(z,t)= \alpha( z+2t \partial)\partial f(z,t)
+\beta(z+2t \partial)\partial_t f(z,t)
\label{3.25}
\eeq
 Eq.\ref{3.20} is straightforwardly extended to $CP^2$ where
it takes the form
\beq
\partial_z\partial_t f^A([z])= L([z]) \partial_t f^A([z])+
M([z]) \partial_z f^A([z]).
\label{3.26}
\eeq
Using this relation, and  Eq.\ref{3.22}, we may re-express all
higher derivetives in Eq.\ref{3.25} in terms of the first-order ones,
obtaining  a transformation of form
\beq
\delta f(z,t)= A( z, t) \partial f(z,t)
+B( z, t)\partial_t f(z,t)
\label{3.27}
\eeq
which correspond to the change of parametrization
\beq
\delta z =A( z, t),\quad \delta t =B( z, t),\>  \hbox{with}\>
\delta f(z,t)=f (z+\delta z, t+\delta t)- f(z,t).
\label{3.28}
\eeq
Thus W transformations are extended as particular diffeomorphisms of
$CP^2$.
\subsection{Light-cone description}
Finally, we connect the above discussion with  the light-cone approach
to $W_3$ gravity, which so far was developed independently
 of the conformal gauge\cite{M1},\cite{LC}. The basic point, displayed
in ref.\cite{GM3}, is that,   for any K\"ahler manifold, there exist
parametrizations such that the metric takes  a  block-form
identical to  the
light-cone metric Eq.\ref{2.21}
 introduced by Polyakov for two-dimensional gravity.
\subsubsection{Light-cone parametrization for K\"ahler
manifolds}
Consider an arbitrary K\"ahler manifold, introduced by Definition 1.
There exists
another class  of prefered
 coordinates denoted $U^A$, and $U^{\Ab}$, such that
the new metric tensor denoted by the letter $H$
takes a form which is  the generalization of the right-hand side
 of Eq.\ref{2.21}. The change of coordinates is of the form
\beq
U^{\Ab}=
U^{\Ab}(X^1,\, \cdots,\, X^n; X^{\1b} \, \cdots,\,  X^{\nb} ), \quad
 U^{A}=X^A.
\label{2}
\eeq
The functions  $U^{\Ab}(X; \Xb)$ will be determined next, in
such a way that
 the light-cone metric takes the form
$$H_{\Ab \, \Bb}=0,  \quad H_{A \, B} =2 h_{A \, B}
$$
\beq
H_{A \, \Bb}= H_{\Bb \, A} =
\delta_{A, \, \Bb},\hfill
\label{3}
\eeq
where $h_{A \, B}$ will be related to the K\"ahler potential.
Before going on, let us remark that the determinant of $H $ is
equal to $-1$, and that
its   inverse  is given by
$$
H^{A \, B}=0, \quad H^{\Ab \, \Bb}=
-2 h_{A \, B}
$$
\beq
H^{A \, \Bb}= H^{\Bb \, A} =
\delta_{A, \, \Bb}.\hfill
\label{4}
\eeq
On the contrary, no general form may be given for the determinant of $G$
or its inverse. This is a very nice point of the
light-cone parametrization\footnote{inverting the metric tensor
is often a pain in the neck.}.
 Going back to our main line, one
finds,
by standard computations,  that  conditions Eqs.\ref{3}
are fulfilled if one has
\beq
G_{A \, \Bb}=H_{A \, \Cb} \partial_{\Bb}U^{\Cb}
\label{5}
\eeq
\beq
2h_{A \, B}=-H_{B \, \Cb} \partial_A U^{\Cb}
-H_{A \, \Cb} \partial_B U^{\Cb}
\label{6}
\eeq
These equations
are  easily solved  by using the K\"ahler potential,
obtaining
\beq
U^{\Ab}=H^{\Ab \, C} \partial_C K,
\label{7}
\eeq
\beq
h_{A \, B}=-\partial_A \partial_B K.
\label{8}
\eeq
Thus we reach the important conclusion that, for any
K\"ahlerian manifold there is a choice of coordinates
such that the metric tensor takes the block-form
\beq
H=\left ( \begin{array}{cc}
2h&1\\
1&0
\end{array} \right )
\label{9}
\eeq
which is the same as the one  introduced by Polyakov
to describe  2D gravity in the light-cone gauge.

\subsubsection{Light-cone coordinates for $W_3$ gravity}
We finally apply this change of coordinates to $CP^2$, calling
$\ub^1$, and $\ub^2$ the two light-cone coordinates. Combining
Eq.\ref{3.20} with Eq.\ref{7}, \ref{8}, one sees that  one has
\beq
\ub^\ell= {\sum_{A=0}^2 \partial_\ell f^A([z]) \fb^A([\zb])\over
\sum_{A=0}^2 f^A([z]) \fb^A([\zb])}
\label{3.37}
\eeq
\[
h_{\ell m}= -{\sum_{A=0}^2 \partial_\ell \partial_m
f^A([z]) \fb^A([\zb])\over
\sum_{A=0}^2 f^A([z]) \fb^A([\zb])} +\ub^\ell \ub^m
\]
Making use of Eq.\ref{3.26}, this gives
\[
h_{11}=- \ub^2+(\ub^1)^2, \quad
h_{12}=-L \ub^2-M\ub^1+ \ub^1\ub^2
\]
\beq
h_{22}= - \ub^2 (\partial L +L^2 +M) -\ub^1 (\partial M +LM) +
(\ub^2)^2
\label{3.38}
\eeq
so that $h$ is quadratic in the $\ub$'s.
On the other hand, Polyakov's equation for
light-cone $W_3$ gravity takes the form $(\Db)^3 h =0$, and
$(\Db)^5 A =0$. This contradiction may be  resolved as follows:
{\bf The W surface where light-cone physics
takes place is not at $z^{(2)} =\zb^{(2)}=0$, contrary to
conformal physics.} Indeed, let us assume that the light-cone W
surface has equations
\beq
z^{(2)}=0,\quad \ub^2=\alpha(z)\,  (\ub^1)^2+\beta(z)\,  \ub^1
+\gamma(z)
\label{3.39}
\eeq
Then, on this surface, $h_{\ell m}$ is a polynomial in
$\Ub^1$ of degree $\ell+m$. Thus it seems natural to let
\beq
h_{11}=h,\quad h_{22}=A
\label{3.40}
\eeq
obtaining
\beq
( \Db)^3 h=0, \quad
(\Db)^5 A=0, \quad \hbox{with} \>  \Db=\Db_1
\label{3.41}
\eeq
Clearly, $\alpha$, $\beta$, and $\gamma$ are arbitrary.
They may be precisely changed by an $sl(3)$ transformation. Thus,
 one
is led to conjecture that {\bf the geometrical origin of the $sl(3)$
invariance is a change of the light-cone W surface so that
Eq.\ref{3.38} keeps its form}.

\section{Outlook}

These lecture notes left out many aspects, some of which have been developed
elsewhere. In particular, quantization has  led to many interesting
progress. For the Liouville case,
it was realized\cite{B,G1}
 that there exists a quantum-group\index{Quantum groups} structure where
the mathematical parameter of the quantum-group deformation coincides
with Planck's constant. In this theory, the non-commutativity
which is inherent
in the quantum group -- since the co-product in non-symmetric --
is brought about by the actual quantization of the system.
 This
appearance of quantum group
seems  to be very natural geometrically, since one deals with a
gravity theory, where the space-time metric is quantized, which seems
tantamount to quantizing the 2D space-time itself. Thus an object like
a quantum plane should appear, and quantum groups
are natural transformations of such ``quantum manifolds''. For recent
progress in this direction, see refs. \cite{CGR1,CGR2,GS2}.
The quantum group
picture was extended to the quantum $A_n$ Toda theories in ref.\cite{CG}.
The classical geometry  setting
summarized in these notes indicates that
at the quantum level  one should be dealing with
  quantum complex projective  spaces. This remains to be clarified.

Classically, what we describe is also incomplete. In the  light-cone
connection, the ansatz Eq.\ref{3.40}  remains  somewhat ad hoc, and its
 Kac-Moody        structure should be clarified.
Since we use the Toda equation
throughout, we stayed on the mass shell.
Thus we only get Polyakov equations (Eqs.\ref{2.25}, \ref{3.41}),
and not the full anomaly equations  that characterize light-cone
W gravities.
Besides the particular change of gauge we have discussed,
the most basic problem is to formulate W gravity without gauge fixing, and
to couple it to some appropriate kind of matter from first principle.

\end{document}